\documentclass[pra,aps,twocolumn]{revtex4}

\usepackage{graphicx,amsmath}
\usepackage{txfonts}
\usepackage{hyperref}
\newcommand{\non}{\nonumber}

\newcommand{\rp}[1]{(\ref{#1})}

\newcommand{\abs}[1]{\left|{#1}\right|}

\newcommand{\av}[1]{\left\langle #1 \right\rangle}

\newcommand{\al}[1]{^{(#1)}}
\newcommand{\da}{^\dagger}

\newcommand{\pt}[1]{\left( #1 \right)}
\newcommand{\pq}[1]{\left[ #1 \right]}

\newcommand{\lpq}[1]{\left[ #1 \right.}
\newcommand{\lpg}[1]{\left\{ #1 \right.}

\newcommand{\rpq}[1]{\left. #1 \right]}
\newcommand{\rpg}[1]{\left. #1 \right\}}
\newcommand{\ee}{{\rm e}}
\newcommand{\ii}{{\rm i}}
\newcommand{\dd}{{\rm d}}

\newcommand{\nn}{{\nonumber}}

\newcommand{\EE}{{\cal E}}

\newcommand{\RR}{{\cal R}}

\begin{document}
\title{Optomechanical cooling with intracavity squeezed light}

\author{Muhammad Asjad,$^1$ Najmeh Etehadi Abari,$^{2,3}$ Stefano Zippilli,$^2$ David Vitali$^{2,4,5}$}
\affiliation{$^1$ Max Planck Institute for the Science of Light, Staudtstra{\ss}e 2, D-91058 Erlangen, Germany}
\affiliation{$^2$ School of Science and Technology, Physics Division, University of Camerino, I-62032 Camerino, Italy}
\affiliation{$^3$ Department of Physics, Faculty of Science, University of Isfahan, Hezar Jerib, 81746-73441, Isfahan, Iran}
\affiliation{$^4$ INFN, Sezione di Perugia, I-06123 Perugia, Italy}
\affiliation{$^5$ CNR-INO, Largo Enrico Fermi 6, I-50125 Firenze, Italy}
\date{\today}

\begin{abstract}
We analyze the performance of optomechanical cooling of a mechanical resonator in the presence of a degenerate optical parametric amplifier within the optomechanical cavity, which squeezes the cavity light. We demonstrate that this allows to significantly enhance the cooling efficiency via the coherent suppression of Stokes scattering. The enhanced cooling occurs also far from the resolved sideband regime,
and we show that this cooling scheme can be more efficient than schemes realized by injecting a squeezed field into the optomechanical cavity.
\end{abstract}
\maketitle

\section{Introduction}

In cavity optomechanics~\cite{Bowen},
squeezed light has been studied
as a powerful tool for improved detection sensitivity~\cite{LIGO,Peano,Korobko,Clark}, for the preparation of quantum states of mechanical resonators~\cite{Jahne,Asjad2016b},
to achieve larger optomechanical coupling strength~\cite{Huang2009b,Lu}, and for enhanced optomechanical cooling~\cite{Huang2009a,Asjad2016a,Clark2017,Lau2019}.
In general one can consider two different scenarios. On the one hand squeezed light produced by an optical parametric oscillator (OPO) external to the optomechanical cavity can be injected into the system, and on the other the OPO can be placed within the optomechanical cavity to directly squeeze the cavity light fluctuations.
In this paper we are interested in optomechanical cooling. In particular, we focus on the scheme with internal OPO, we study under which conditions it is able to outperform standard sideband cooling techniques, and we analyze in details how it compares with the results achievable in the case of injected
squeezing.

The cooling dynamics in the case of injected squeezing has been analyzed~\cite{Asjad2016a,Clark2017} and experimentally demonstrated~\cite{Clark2017}. It has been shown that squeezed light can be tuned to coherently suppress the scattering Stokes process heating the resonator, hence allowing to surpass the quantum backaction limit of optomechanical sideband cooling~\cite{Aspelmeyer2014} (see also Refs.~\cite{Kralj2017,Rossi2017,Zippilli2018} for an alternative, feedback-based method for beating the quantum backaction limit).
Here we show that, as far as the cavity linewidth is not very large,
the optimal cooling achieved with a degenerate OPO placed within the optomechanical cavity~\cite{Huang2009a}  is due to a similar destructive interference which contributes to the coherent suppression of the dominant Stokes scattering processes.
In fact in this regime, the cooling efficiency of the two schemes is the same, and the two models can be mapped one into the other by a unitary transformation
(see also Ref.~\cite{Lau2019}), even though the optimal performance is achieved at different sets of parameters. In particular the cooling with intracavity squeezing is optimized for larger squeezing.
\begin{figure}[th!]
\centering
 \includegraphics[width=.3\textwidth]{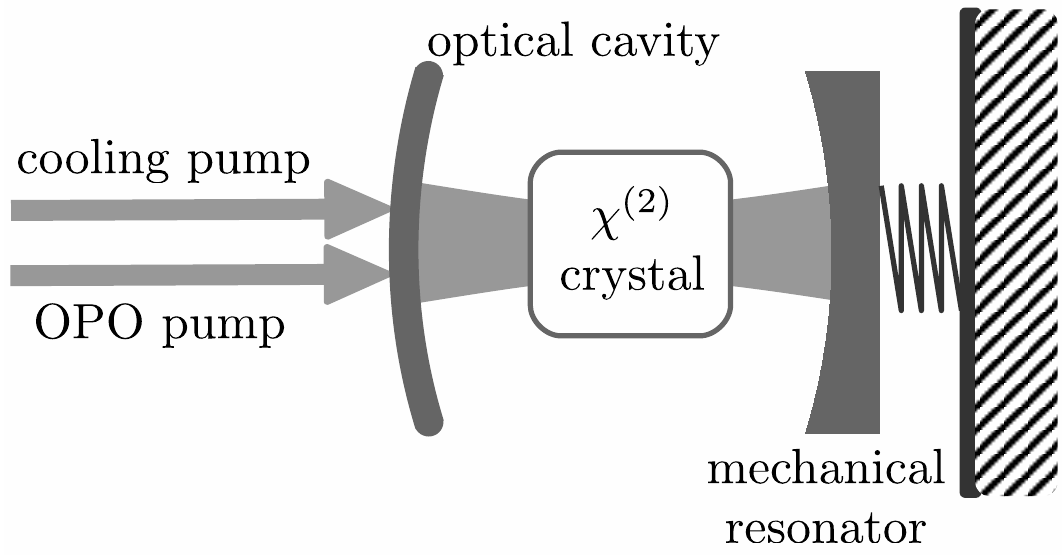}
 \caption{Schematical description of the optomechanical system with intracavity squeezing generated by the driven nonlinear crystal.}\label{fig0}
 \end{figure}

As observed in Ref.~\cite{Asjad2016a,Lau2019}, the cooling dynamics with
squeezed light (both injected or with internal OPO) are very efficient also in the unresolved sideband regime, that is when the cavity linewidth is much larger than the mechanical frequency. In this case, however, the two schemes are no more equivalent. Specifically, the
scheme with the internal OPO results more efficient, and its optimal performance is achieved when the OPO is driven close to the optical instability threshold.
Finally, we also study the mechanical effect of the OPO pump and show that in general it tends to reduce the cooling efficiency.

The article is organized as follows. In section~\ref{Sec:model} we introduce the system model. In section~\ref{Results} we analyze the cooling dynamics in the limit in which the mechanical effects of the OPO pump fields can be neglected. In particular we study the regime in which the models with injected and intracavity squeezing are equivalent and we identify the parameters for which the cooling using the internal OPO is superior. Then, in section~\ref{Sec:OPOpump} we study the effect of the OPO pump field. Finally section~\ref{conclusions} is for the conclusions.

\section{Model}\label{Sec:model}

We consider an optical parametric oscillator (OPO) consisting of a Fabry-Perot cavity  with a second order non-linear medium (see Fig.~\ref{fig0}) which down-converts photons of a cavity mode at frequency $\omega_c$ to photons at frequency $\omega_a=\omega_c/2$ which match a second resonant mode of the cavity according to the Hamiltonian
\begin{eqnarray}
H_{OPO}=\ii\hbar \frac{\chi_0}{2}\pt{c\ {a\da}^2-c\da \ {a}^2},
\end{eqnarray}
where $a$, $a\da$, $c$ and $c\da$ are the bosonic operators for the annihilation and creation of photons of the two cavity modes.
One cavity mirror is movable so that the resonant cavity frequencies result modulated by the mechanical vibrations. A single vibrational mode, at frequency $\omega_m$ and with annihilation and creation operators $b$ and $b\da$ is assumed relevant, so that the optomechanical interaction Hamiltonian is given by
\begin{eqnarray}
H_{OM}=-\hbar\ g_a\  a\da a \pt{b+b\da}-\hbar\ g_c\ c\da c \pt{b+b\da}\ .
\end{eqnarray}
Both cavity modes are driven by quasi resonant laser fields at frequencies $\omega_{L,a}$ and $\omega_{L,c}$ detuned by $\bar\Delta_\iota=\omega_\iota-\omega_{L,\iota}$ (for $\iota=a,c$) from the corresponding cavity frequencies, and with strength $\EE_\iota$, so that, including optical and mechanical dissipation at rates $\kappa_a$, $\kappa_c$ and $\gamma$,
the corresponding quantum Langevin equations are given by
\begin{eqnarray}\label{Eq0}
\dot a&=&-\pt{\kappa_a+\ii\,\bar\Delta_a}a+\ii\,g_a\ a\pt{b+b\da}+\chi_0\,c\,a\da+\EE_a+\sqrt{2\kappa_a}\,a_{in},
\nn\\
\dot c&=&-\pt{\kappa_c+\ii\,\bar\Delta_c}c+\ii\,g_c\ c\pt{b+b\da}-\frac{\chi_0}{2}\,a^2+\EE_c+\sqrt{2\kappa_c}\,c_{in},
\nn\\
\dot b&=&-\pt{\frac{\gamma}{2}+\ii\,\omega_m}b+\ii\,g_a\ a\da a+\ii\,g_c\ c\da c+\sqrt{\gamma}\,b_{in},
\end{eqnarray}
where $a_{in}$, $c_{in}$ and $b_{in}$ are the input noise operators. $a_{in}$ and  $c_{in}$ describe vacuum noise with correlations $\av{a_{in}(t)\, a_{in}\da(t')}=\av{c_{in}(t)\, c_{in}\da(t')}=\delta(t-t')$, while $b_{in}$ describes thermal noise for the mechanical resonator and is characterized by the correlation function $\av{b_{in}(t)\,b_{in}\da(t')}=(n_T+1)\delta(t-t')$ with $n_T=\pt{\ee^{\hbar \omega_{m}/K_B T}-1}^{-1}$
the mean thermal occupation number of the mechanical mode at temperature $T$.

When the pump fields are strong, the system equation in Eq.~\rp{Eq0} can be linearized around the steady state average values of the
optical and mechanical variables ($\av{a}_{st}$, $\av{b}_{st}$ and $\av{c}_{st}$) neglecting non-linear terms in the fluctuations of the system operators $\delta \iota=\iota-\av{\iota}_{st}$ for $\iota=a,b,c$. The parameters $\av{a}_{st}$, $\av{b}_{st}$ and $\av{c}_{st}$
fulfill the relations
\begin{eqnarray}
 - \pq{ \kappa_{a}+ \ii\,\Delta_{a} }\av{a}_{st}+ \chi_0\, \av{a}^*_{st} \av{c}_{st}+\EE_{a}  &=& 0,\nn\\
- \pq{\kappa_{c}+ \ii\,\Delta_{c} }\av{c}_{st} - \frac{\chi_0}{2}\av{a}_{st}^{2}+\EE_{c}  &=& 0,\nn\\
- \pt{\frac{\gamma}{2} + \ii\,  \omega_{m}}\av{b}_{st}+
\ii \pt{ g_{a} \abs{\av{c}_{st}}^2 + g_{a} \abs{\av{a}_{st}}^2 } &=& 0.
\end{eqnarray}
where $\Delta_\iota=\bar\Delta_\iota-g_\iota\av{b+b\da}$, for $\iota=a,c$,
that is,
\begin{eqnarray}
\av{a}_{st}&=&\dfrac{\EE_a\pt{\kappa_a-\ii\Delta_a}+\EE_a^*\ \chi_0 \av{c}_{st}}{\kappa^2_a+\Delta^2_a-\chi_0^2 \abs{\av{c}_{st}}^2},
\nn\\
\av{c}_{st}&=&\dfrac{\EE_c}{\kappa_c+\ii\Delta_c}
-\frac{\chi_0}{2}\dfrac{\av{a}_{st}^{2}}{\kappa_c+\ii\Delta_c}
,\non\\
\av{b}_{st}&=&\ii\dfrac{g_a\abs{\av{a}_{st}}^2 +g_c\abs{\av{c}_{st}}^2}{{\frac{\gamma}{2}}+\ii\omega_m}\ ,
\end{eqnarray}
with $\Delta_\iota=\bar\Delta_\iota-g_\iota\left[\av{b}_{st}+\av{b}_{st}^{*}\right]$, for $\iota=a,c$.
The corresponding equations for the fluctuations are
\begin{eqnarray}
\dot{\delta a}&=&-\pt{\kappa_a+\ii\,\Delta_a}{\delta a}+\ii\,G_a\pt{\delta b+\delta b\da}+\chi\,\delta a\da+\epsilon^*\,\delta c
\nn\\&&
+\sqrt{2\kappa_a}\,a_{in}\ ,
\nn\\
\dot{\delta c}&=&-\pt{\kappa_c+\ii\,\Delta_c}\delta c+\ii\,G_c\pt{\delta b+\delta b\da}-\epsilon\,\delta a+\sqrt{2\kappa_c}\,c_{in}\ ,
\nn\\
\dot{\delta b}&=&-\pt{{\frac{\gamma}{2}}+\ii\,\omega_m}\delta b+\ii\pt{G_a\,\delta a\da+G_a^*\,\delta a}
\nn\\&&
+\ii\pt{G_c\, \delta c\da+G_c^*\,\delta c}+\sqrt{\gamma}\,b_{in}\ ,
\end{eqnarray}
where
\begin{eqnarray}
G_\iota=g_\iota\av{\iota}_{st} \hspace{1cm}{\rm for}\ \  \iota=a,c
\end{eqnarray}
is the linearized optomechanical coupling,
\begin{eqnarray}
\chi=\chi_0\av{c}_{st}
\end{eqnarray}
is the strength of the non-linear self-interaction of the mode $a$, and
\begin{eqnarray}
\epsilon=\chi_0\av{a}_{st}
\end{eqnarray}
is the strength of the effective beam splitter interaction between $\delta a$ and $\delta c$.

Without loss of generality we assume the phase of $\av{a}_{st}$ to be zero so that $G_a$ and $\epsilon$ are real, moreover we explicitly define the phase of $\av{c}_{st}=\abs{\av{c}_{st}}\ \ee^{2\ii\phi}$, and we use the substitution $G_c\to G_c\,\ee^{2\ii\phi}$ and $\chi\to\chi\,\ee^{2\ii\phi}$ (with the new $G_c$ and $\chi$ real), such that
\begin{eqnarray}\label{OPO1}
\dot{\delta a}&=&-\pt{\kappa_a+\ii\,\Delta_a}{\delta a}+\ii\,G_a\pt{\delta b+\delta b\da}+\chi\,\ee^{2\ii\phi}\,\delta a\da+\epsilon\,\delta c
\nn\\&&
+\sqrt{2\kappa_a}\,a_{in},
\nn\\
\dot{\delta c}&=&-\pt{\kappa_c+\ii\,\Delta_c}\delta c+\ii\,G_c\,\ee^{2\ii\phi}\pt{\delta b+\delta b\da}-\epsilon\,\delta a+\sqrt{2\kappa_c}\,c_{in},
\nn\\
\dot{\delta b}&=&-\pt{{\frac{\gamma}{2}}+\ii\,\omega_m}\delta b+\ii\,G_a\pt{\delta a\da+\delta a}
\nn\\&&
+\ii\,G_c\pt{\ee^{2\ii\phi}\, \delta c\da+\ee^{-2\ii\phi}\,\delta c}+\sqrt{\gamma}\,b_{in}\ .
\end{eqnarray}
We remark that although
we have described a Fabry-Perot configuration, this model remains valid for any geometry of the optical and mechanical system. In particular promising devices which can support the dynamics discussed hereafter are micro--resonators which sustain both optical and mechanical modes~\cite{Hofer}, and which could be realized with a
non-linear material able to provide the parametric amplification~\cite{Furst,Otterpohl}.

\subsection{Cooling}\label{Sec:cooling}

The cavity light can be used to cool the mechanical vibrations.
Light acts as an effective thermal reservoir which compete with the natural thermal reservoir to determine the stationary mechanical energy. The effective light-controlled reservoir is characterized by an effective number of mechanical excitations $n_o$ (to which the resonator would thermalize in the absence of the natural thermal environment) and by a rate $\Gamma$ at which mechanical excitations are dissipated. Thereby the stationary number of mechanical excitations is given by
\begin{eqnarray}\label{Nst}
N_{st}=\frac{\gamma\ n_T+\Gamma\ n_o}{\gamma+\Gamma}\ .
\end{eqnarray}
The parameters $n_o$ and $\Gamma$ are determined by how light is scattered by the mechanical resonator. Specifically
the resonator can scatter inelastically incident photons to lower (Stokes processes) and higher (anti-Stokes processes) frequencies when at the same time mechanical energy is increased and reduced respectively. The rates for Stokes ($A_+$) and anti-Stokes ($A_-$) processes are determined by the fluctuations of the cavity field. In particular in the limit of weak optomechanical coupling they can be evaluated perturbatively at the lowest relevant order in terms of the power spectrum of the field operator which is coupled to the mechanical resonator (the power spectrum of the radiation pressure force operator), in our case
\begin{eqnarray}\label{F}
F=- \sum_{\iota=a,c} G_\iota\pt{\delta \iota+\delta \iota\da}\ ,
\end{eqnarray}
evaluated at $\pm\omega_m$. Specifically,
\begin{eqnarray}\label{fullApm}
A_\pm=\int_{-\infty}^\infty\,\dd t\ \ee^{\mp\ii\omega_m\,t}\av{F(t)\ F(0)}_{st},
\end{eqnarray}
where the average is evaluated at the steady state of an empty cavity (without the mechanical resonator).
In terms of these scattering rate one finds
\begin{eqnarray}
\Gamma&=&{A_--A_+}\ ,
\nn\\
n_o&=&\frac{A_+}{\Gamma}\ ,
\end{eqnarray}
which are valid when
cooling takes places, that is when $A_->A_+$.

\begin{figure*}[th!]
\centering
\includegraphics[width=.85\textwidth]{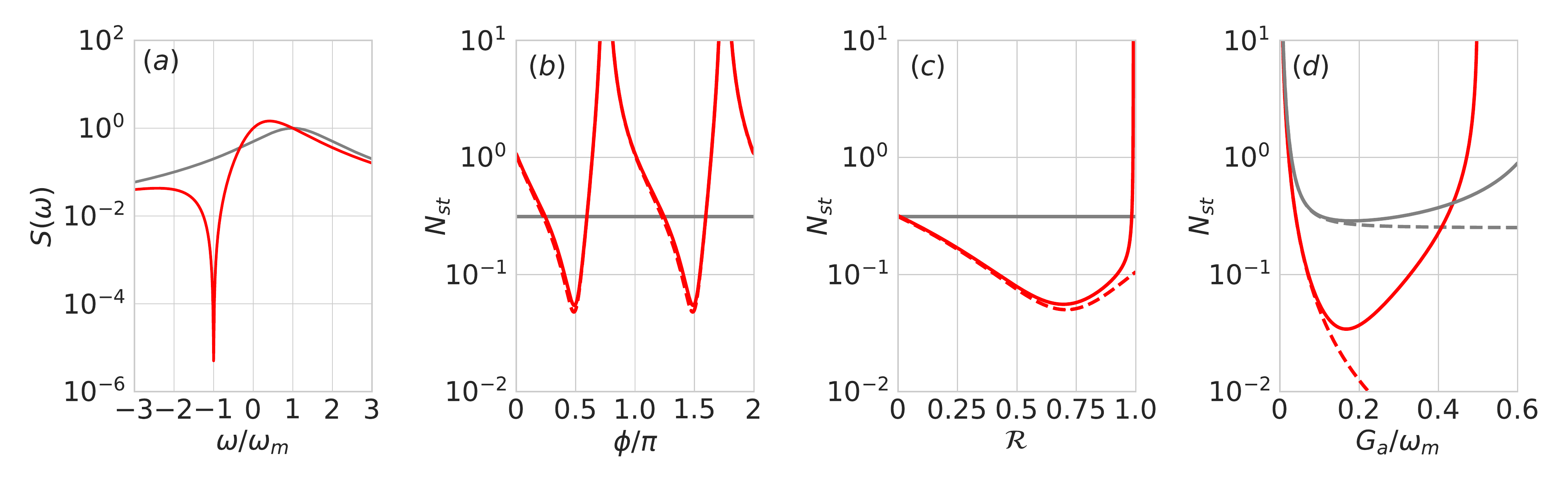}
\caption{
(a) Power spectrum of the radiation pressure force defined in Eq.~\rp{Sa} and (b)-(d) steady-state excitation number $N_{st}$ as a function of (b) the phase of the squeezing $\phi$, (c) the value of squeezing measured by the parameter $\RR$ defined in Eq.~\rp{Rchi}, and (d) the optomechanical coupling strength $G_a$.
The red solid lines are evaluated by solving numerically Eq.~\rp{OPO0}. The red dashed lines correspond to the analytical approximated solution evaluated with Eq.~\rp{Sa}. Solid and dashed grey lines are numerical and analytical results evaluated without squeezing, namely for $\chi=0$ (i.e $\RR=0$).
The red lines are evaluated with the optimal values for Stokes scattering suppression $\phi\al{opt}$ and $\mathcal{R}=\RR\al{opt}$ as reported in Eqs.~\rp{optOPO} and \rp{oprR}.
The other parameters are $n_T=1000$, $\gamma_m=0.25\times10^{-6}\omega_m$ and $\kappa_a= \omega_m$, $G_a=0.1\omega_m$, $\Delta_a=\omega_m$.
}
\label{fig1}
\end{figure*}

\section{Injected vs intracavity squeezing}\label{Results}

Let us first study the limit
of large $\abs{\kappa_c+\ii\,\Delta_c}$. In this limit the effect of the fluctuations of mode $c$ can be neglected and the system equations reduce to
\begin{eqnarray}\label{OPO0}
\dot{\delta a}&=&-\pt{\kappa_a+\ii\,\Delta_a}{\delta a}+\ii\,G_a\pt{\delta b+\delta b\da}+\chi\,\ee^{2\ii\phi}\,\delta a\da
+\sqrt{2\kappa_a}\,a_{in},
\nn\\
\dot{\delta b}&=&-\pt{{\frac{\gamma}{2}}+\ii\,\omega_m}\delta b+\ii\,G_a\pt{\delta a\da+\delta a}
+\sqrt{\gamma}\,b_{in}\ .
\end{eqnarray}
We provide a detailed justification of this approximation in Sec.~\ref{Sec:OPOpump}, where we also  show that, when the system is not well within this limit, the effects of $\delta c$ is, in general, to reduce the cooling efficiency.

When the nonlinearity is not too strong $\chi<\abs{\Delta_a}$, the model described by Eq.~\rp{OPO0} is equivalent to a model in which a standard optomechanical system is driven by a squeezed reservoir (see also Ref. \cite{Lau2019}). This is evident when we study the dynamics for a squeezed cavity operator of the form
\begin{eqnarray}\label{as}
\delta a_s=\ee^{\ii\,\phi'}\pq{\cosh\pt{s}\ \delta a+\sinh\pt{s}\ \ee^{2\ii\,\pt{\phi_s-\phi'}}\ \delta a\da }\ ,
\end{eqnarray}
with
\begin{eqnarray}\label{tanhs}
\tanh(s)&=&\frac{\Delta_a-\sqrt{\Delta_a^2-\chi^2}}{\chi},
\\
\phi_s&=&\phi+\phi'+\frac{\pi}{4},
\\
\phi'&=&\arctan\pq{\frac{\sinh(s)\ \sin\pt{2\phi_s}}{\cosh(s)+\sinh(s)\ \cos\pt{2\phi_s}}}\ .
\end{eqnarray}
We find that the corresponding quantum Langevin equations read
\begin{eqnarray}
\label{SQR}
\dot{\delta a_s}&=&-\pt{\kappa_a+\ii\,\Delta_a\al{s}}{\delta a_s}+\ii\,G_a\al{s}\pt{\delta b+\delta b\da}
+\sqrt{2\kappa_a}\,a_{in}\al{s},
\nn\\
\dot{\delta b}&=&-\pt{{\frac{\gamma}{2}}+\ii\,\omega_m}\delta b+\ii\,G_a\al{s}\pt{\delta a_s\da+\delta a_s}
+\sqrt{\gamma}\,b_{in},
\end{eqnarray}
with modified parameters
\begin{eqnarray}\label{chi2s}
\Delta_a\al{s}&=&\frac{\Delta_a}{2\,n_s+1},
\nn\\
G_a\al{s}&=&G_a\sqrt{\frac{n_s}{m_s^2+n_s^2+2\,n_s\,m_s\ \cos\pt{2\phi_s}}}\ ,
\end{eqnarray}
where
\begin{eqnarray}
n_s&=&\sinh(s)^2,
\nn\\
m_s&=&\cosh(s)\ \sinh(s)=\sqrt{n_s(n_s+1)}\ .
\end{eqnarray}
The new input noise for the cavity field is
characterized by squeezed fluctuations, such that, in the frequency domain,
$\av{{a_{in}\al{s}}\da(\omega)\ a_{in}\al{s}(\omega')}=
\delta(\omega+\omega')\ n_s$
and
$\av{a_{in}\al{s}(\omega)\ a_{in}\al{s}(\omega')}=\delta(\omega+\omega')\ m_s\, \ee^{-2\ii\phi_s}$
where
the coefficients $n_s$ and $m_s$ determine respectively the number of excitations and the strength of the correlations of the input squeezed field.

The equations in the new representation, Eq.~\rp{SQR}, are equivalent to the equations for a system driven by the output field of an external parametric amplifier (as the one investigated in Refs.~\cite{Asjad2016a,Clark2017}) in the broadband limit, and with no losses in the transmission and injection of the squeezed field form the external OPO into the optomechanical system. We therefore expect to observe the same dynamics described in Refs.~\cite{Asjad2016a,Clark2017} with the present system. In particular also in this case we expect that Stokes scattering processes can be suppressed so that the backaction limit of standard sideband cooling can be surpassed. It is, however, interesting to study under which parameter regime one observes these dynamics in the present case, and how this compares with the case of squeezing injection from an external OPO.

We further note that the equivalence between the two models [Eqs.~\rp{OPO0} and \rp{SQR}] is true only when $\chi<\abs{\Delta_a}$. When $\chi>\abs{\Delta_a}$ (note however that in any case $\chi$ have to be smaller than $\sqrt{\kappa_a^2+\Delta_a^2}$ in order to guarantee the stability of the OPO) the nonlinear self-interaction term
cannot be set to zero for any transformation of the form of Eq.~\rp{as}.
In the resolved sideband regime optimal cooling is obtained for relatively small $\chi$, hence in a regime in which
the two models are equivalent (see Sec.~\ref{cool1}).
In the unresolved sideband regime, instead, optimal cooling can be obtained for large $\chi$ and specifically when $\sqrt{\kappa_a^2+\Delta_a^2} >\chi>\abs{\Delta_a}$ (see Sec.~\ref{cool2}). In this case the two models are not equivalent, and in Sec.~\ref{cool2} we show that the model with the intracavity squeezing generated by the OPO can be more efficient, especially when the cavity linewidth is large and the system is far from the resolved sideband regime.

\subsection{Stokes and anti-Stokes scattering rates}\label{secStokes}

As discussed in Sec.~\ref{Sec:cooling},
the cooling dynamics can be efficiently analyzed analytically in perturbation theory at the lowest relevant order in the optomechanical couplings.
In this case, only the cavity mode $a$ contributes to the cooling dynamics, and the  heating and cooling [$A_{a,\pm}=G_a\,S_a(\mp\omega_m)$] rates
are given in terms of
the power spectrum $S_a(\omega)=\int_{-\infty}^\infty\,\dd t\ \ee^{\ii\omega\,t}\av{X_a(t)\ X_a(0)}_{st}$, with
$X_a=\delta a+\delta a\da$,
which can be computed by solving Eqs.~\rp{OPO0} for $G_a=0$. The explicit result is
\begin{eqnarray}\label{Sa}
S_{a}(\omega)=2\ \kappa_a\ \abs{\frac{\nu(\omega)}{\zeta(\omega)}}^2\ ,
\end{eqnarray}
where
\begin{eqnarray}\label{nuzeta}
\nu(\omega)&=&\kappa_a+\ii\pt{\Delta_a+\omega}+\chi\ \ee^{2\ii\phi}\ ,
\nn\\
\zeta(\omega)&=&\pt{\kappa_a-\ii\,\omega}^2+\Delta_a^2-\chi^2\ .
\end{eqnarray}
Similarly we can compute the spectrum by solving Eq.~\rp{SQR} and we find that in the case of external OPO, and in agreement with the findings of Refs.~\cite{Asjad2016a,Clark2017},
\begin{eqnarray}\label{Sas}
S_{a}\al{s}\pt{\omega}&=&2\ \kappa_a\ G_a\al{s}
\nn\\ &&\hspace{-1cm}\times
\abs{
\frac{\sqrt{1+n_s}}{\kappa_a+\ii\pt{\Delta_a\al{s}-\omega}}
+
\frac{\sqrt{n_s}}{\kappa_a-\ii\pt{\Delta_a\al{s}+\omega}}\ \ee^{2\ii\phi_s}
}^2\ .
\end{eqnarray}

 \begin{figure*}[t!]
\centering
 \includegraphics[width=.85\textwidth]{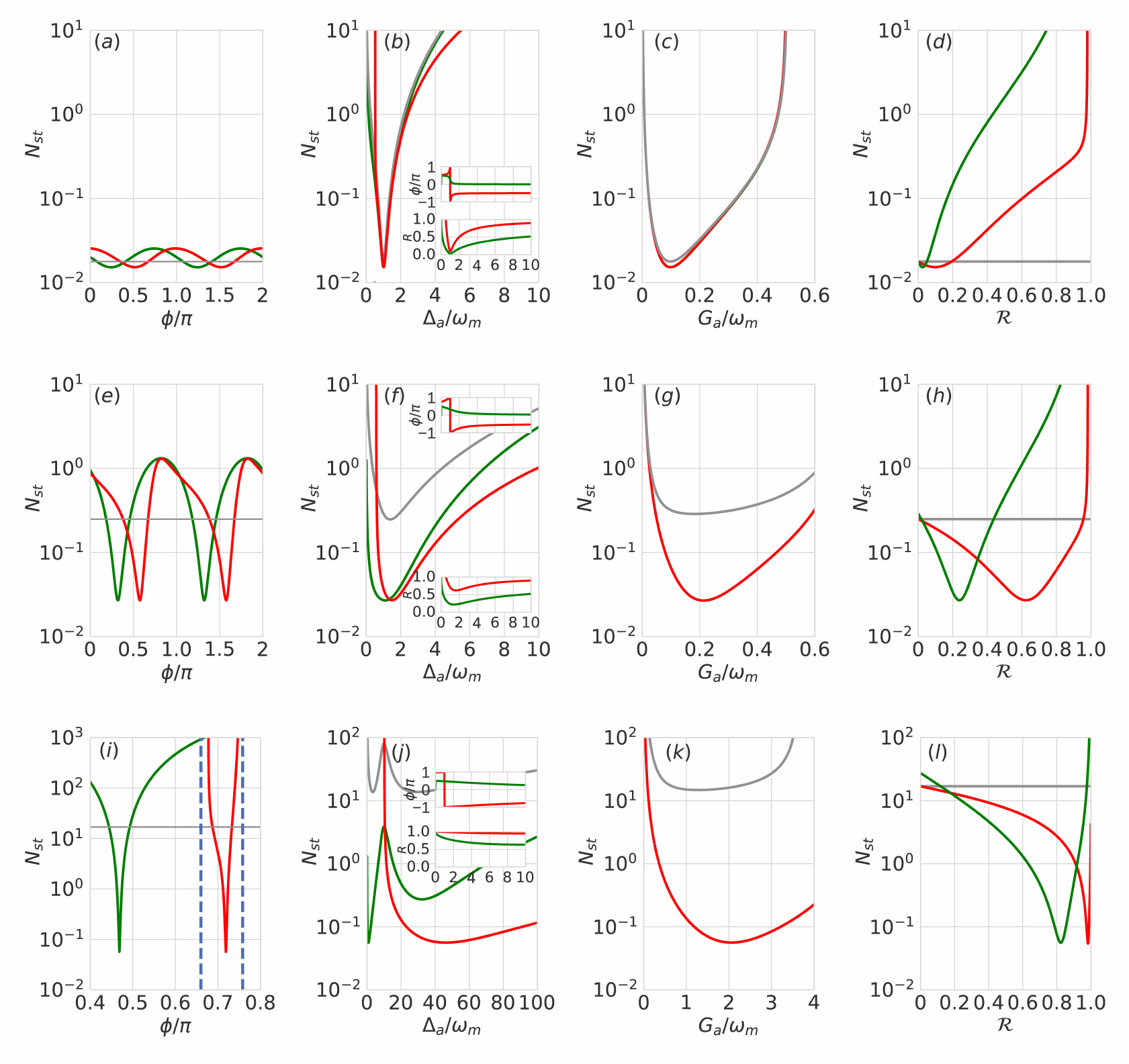}\\
 \caption{
 Steady state excitation number as a function of (first column) the squeezing phase $\phi=\phi_s$, (second column) the optical detuning $\Delta_a=\Delta_a\al{s}$, (third column) the optomechanical coupling strength $G_a=G_a\al{s}$, and (fourth column) the value of squeezing $\RR=\RR_s$.
Each row is evaluated for a different cavity linewidth: (first row) $\kappa_a=0.1\omega_m$, (second row) $\kappa_a=\omega_m$, (third row) $\kappa_a=10\omega_m$. Red lines correspond to the model with internal OPO [see Eq.~\rp{OPO0}], the green lines to the model with external OPO [see Eq.~\rp{SQR}] and the grey lines are with no squeezing.  The vertical dashed lines in plot (i) limit the region of stability of the model described by Eq.~\rp{OPO0}.
The plots in the second and third columns are evaluated under the condition of Stokes scattering suppression defined in Eqs.~\rp{optOPO}, \rp{optSQ}, \rp{oprR} and \rp{optRs}. Each curve is evaluated for the parameters that minimize the corresponding curves in the other plots of the same row.
The insets in the second columns show the specific values of the squeezing parameter $\RR$ and of the squeezing phase $\phi$ and $\phi_s$ used to compute the corresponding results for $N_{st}$. The results for the two models (red and green lines) in the third column are equal and only the red line is visible.
The other parameters are as in Fig.~\rp{fig1}.
  }\label{fig2}
 \end{figure*}

  \begin{figure*}[th!]
\centering
 \includegraphics[width=.85\textwidth]{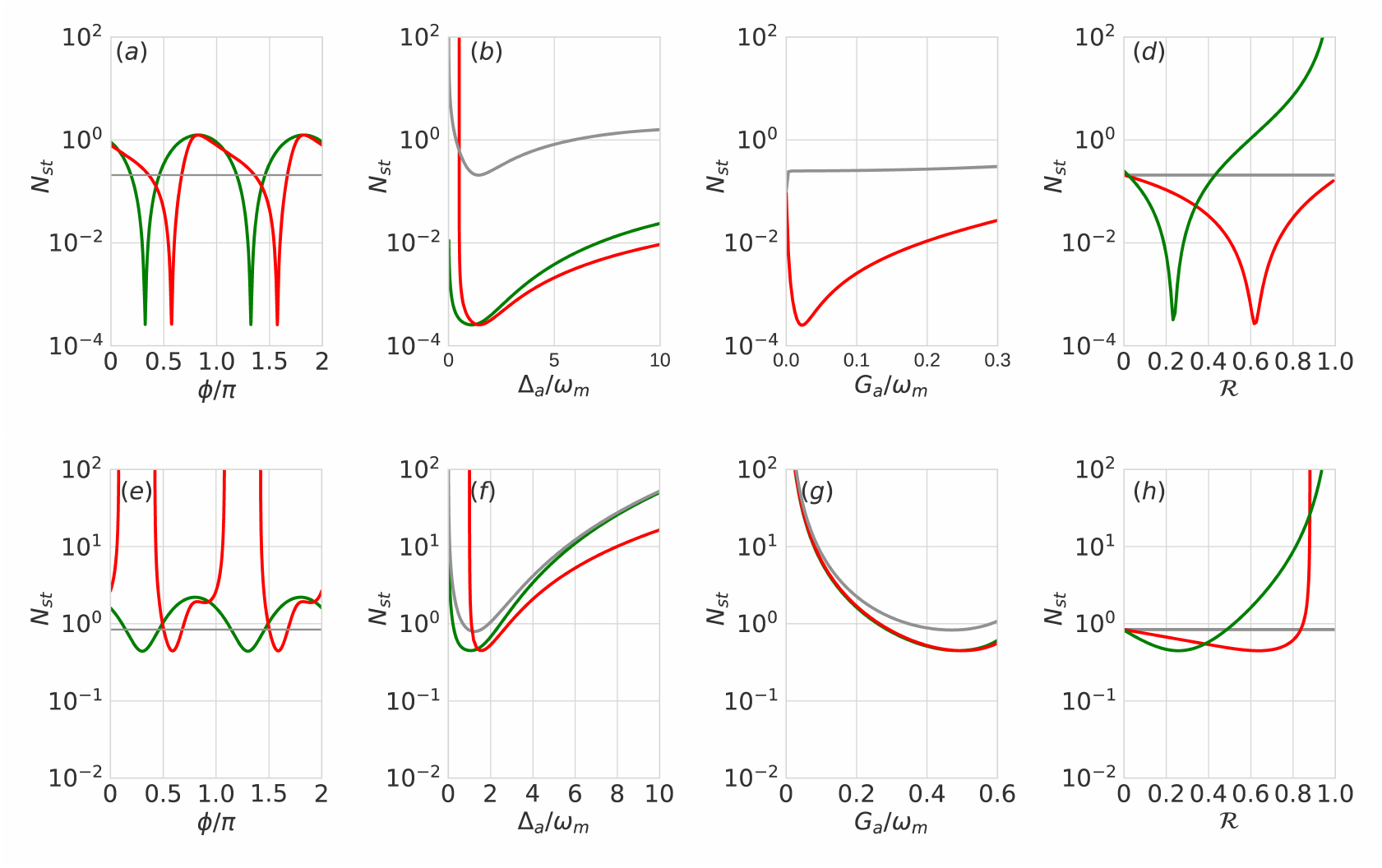}\\
 \caption{Effect of the mechanical reservoir temperature $T$.
The parameters are the same as in the second row of Fig.~\ref{fig2} ($\kappa_a=\omega_m$), but different mean thermal phonon number $n_T$. 
The first row corresponds to $n_T=0.1$ and the second to $n_T=10^5$.
 }\label{fig3}
 \end{figure*}

\subsection{Suppression of Stokes scattering}

We observe that, in the case of internal OPO, Stokes heating processes can be fully suppressed (i.e. $A_{a,+}=0$) when
$\phi=\phi\al{opt}$
and $\chi=\chi\al{opt}$ with
\begin{eqnarray}\label{optOPO}
\ee^{2\ii\phi\al{opt}}&=&-\frac{\kappa_a+\ii\pt{\Delta_a-\omega_m}}{\sqrt{\kappa^2+\pt{\Delta_a-\omega_m}^2}}
\nn\\
\chi\al{opt}&=&\sqrt{\kappa^2+\pt{\Delta_a-\omega_m}^2}\ .
\end{eqnarray}
Similarly, when the squeezing is injected into the optomechanical cavity from an external OPO,
Stokes scattering is suppressed (i.e. $A_{a,+}\al{s}=G_a\al{s}\,S_a\al{s}(-\omega_m)=0$) if~\cite{Asjad2016a,Clark2017}
$\phi_s=\phi_s\al{opt}$
and $n_s=n_s\al{opt}$ with
\begin{eqnarray}\label{optSQ}
\ee^{2\ii\phi_s\al{opt}}&=&-\frac{\pt{\kappa_a-\ii\,\Delta_a\al{s}}^2+\omega_m^2}{\sqrt{\pq{\kappa_a^2+\pt{\Delta_a\al{s}-\omega_m}^2}\pq{\kappa_a^2+\pt{\Delta_a\al{s}+\omega_m}^2}}},
\nn\\
n_s\al{opt}&=&\frac{\kappa_a^2+\pt{\Delta_a\al{s}-\omega_m}^2}{4\,\Delta_a\al{s}\,\omega_m}\ .
\end{eqnarray}
The suppression of Stokes scattering in the case of internal squeezing
is described by Figs.~\ref{fig1}.
These results are very similar to those analyzed in Ref.~\cite{Asjad2016a}.
The spectrum $S_a(\omega)$ is reported in Figs. ~\ref{fig1} (a).
We observe a dip at $\omega=-\omega_m$ which indicates that the Stokes heating processes are strongly suppressed. Correspondingly the cooling is enhanced as described in the other plots, where we report the steady state phonon number, and we compare these results with those of standard sideband cooling, reported as grey lines.
The dip in the spectrum is found for the specific values of the phase and the squeezing determined by Eqs.~\rp{optOPO}.
Plots (b) and (c) describe how the cooling efficiency varies as we move from this optimal condition.
In particular plot (b) reports the result as a function of the phase $\phi$, while plot (c) shows the phonon number as a function of the parameter
\begin{eqnarray}\label{Rchi}
\RR=\frac{\chi}{\sqrt{\kappa_a^2+\Delta_a^2}},
\end{eqnarray}
that determines the amount of squeezing produced by the OPO. In fact,
the variance of the squeezed quadrature ($X_a$) of the cavity field of an OPO (described by Eq.~\rp{OPO0} with $G_a=0$) is $\av{X_a^2}=1/\pt{1+\RR}$ with the value of $\RR$
bounded between $0$ and $1$, with $\RR=1$ indicating the OPO threshold, where the model becomes unstable.
In terms of this parameter the condition for Stokes scattering suppression reads $\RR=\RR\al{opt}$ with
\begin{eqnarray}\label{oprR}
\RR\al{opt}&=&\sqrt{\frac{\kappa^2+\pt{\Delta_a-\omega_m}^2}{\kappa_a^2+\Delta_a^2}}\ .
\end{eqnarray}
In Fig.~\ref{fig1} we also compare the analytical results (dashed lines) obtained with Eq.~\rp{Sa}
to the numerical ones (solid lines) found by the integration of Eq.~\rp{OPO0}.
In particular, plots (b) and (c) are found in a parameter regime in which the value of the optomechanical coupling is sufficiently small so that the approximated analytical results presented above are valid. Deviations between solid and dashed lines are observed in plot (d) for large values of $G_a$.

Let us now anticipate that,
in the following sections,
we compare the cooling efficiency of this model (OPO inside the optomechanical cavity) with the model described by Eqs.~\rp{SQR} where the OPO is external and the squeezed field is injected into the optomechanical cavity.
In particular we
analyze the performance of the two protocols also at fixed source of squeezing. Namely, in both cases
we define a parameter $\RR$ which determines the amount of squeezing produced by the corresponding OPO.
In the case of internal squeezing we consider the definition in Eq.~\rp{Rchi}. Instead,
in the case of external squeezing, we consider an external OPO with strength of the non-linearity $\chi_s$ and cavity decay rate $\kappa_s$ that is operated at resonance (as in the case of Ref.~\cite{Asjad2016a}) and we introduce
\begin{eqnarray}
\RR_s=\frac{\chi_s}{\kappa_s}\ ,
\end{eqnarray}
and correspondingly $n_s={4\,\RR_s^2}/{\pt{R_s^2-1}^2}$.
By means of these relations we can express the system equations
in terms of
$\RR_s$ and we can compare the results of the two models for $\RR=\RR_s$.
In particular, the conditions for the suppression of Stokes scattering,
in the case of external squeezing, can be expressed as
$\RR_s=\RR_s\al{opt}$ with
\begin{eqnarray}\label{optRs}
\RR_s\al{opt}&=&\frac{\sqrt{\kappa_a^2+\pt{\Delta_a\al{s}+\omega_m}^2}-2\sqrt{\Delta_a\al{s}\ \omega_m}}{\sqrt{\kappa_a^2+\pt{\Delta_a\al{s}-\omega_m}^2}}\ .
\end{eqnarray}

\begin{figure*}[th!]
\centering
\includegraphics[width=\textwidth]{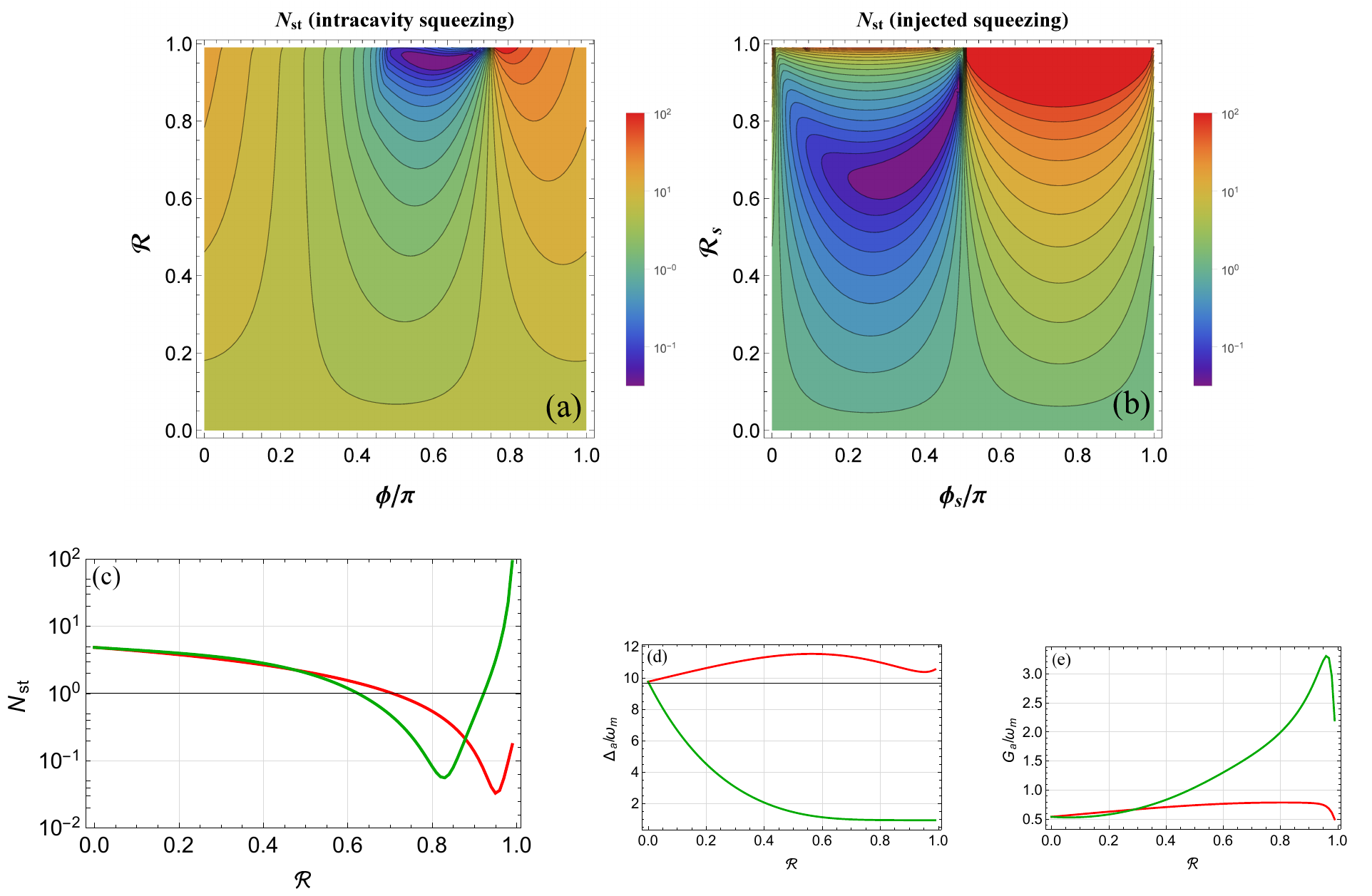}
 \caption{
(a), (b) Steady state excitation number $N_{st}$ as a function of the squeezing phase $\phi$ ($\phi_s$) and the value of squeezing $\RR$ ($\RR_s$), for (a) internal and (b) external OPO, and for $\kappa_a=10\omega_m$.
The curves in plots (c) report the values of $N_{st}$
as a function of $\RR=\RR_s$ along cuts of the contour plots at the specific values of $\phi$ ($\phi_s$) which correspond to the minima of $N_{st}$:
the red curves correspond to the cut of the plots (a) at the value of the phase $\phi=0.62\pi $, and
the green curves to the cut of the plots (b) at the value of the phase  $\phi_s=0.47\pi$.
These results are evaluated by minimizing the value of $N_{st}$, at each point in the $\phi-\RR$ ($\phi_s-\RR_s$) space, as a function of $\Delta_a$ and $G_a$ ($\Delta_a\al{s}$ and $G_a\al{s}$). The specific values of $\Delta_a$ and $G_a$ corresponding to plot (c) are reported in plots (d)-(e).
The other parameters are equal to the ones used in the third row of Fig.\rp{fig2}.
}\label{fig4}
\end{figure*}

\begin{figure*}[th!]
\centering
\includegraphics[width=\textwidth]{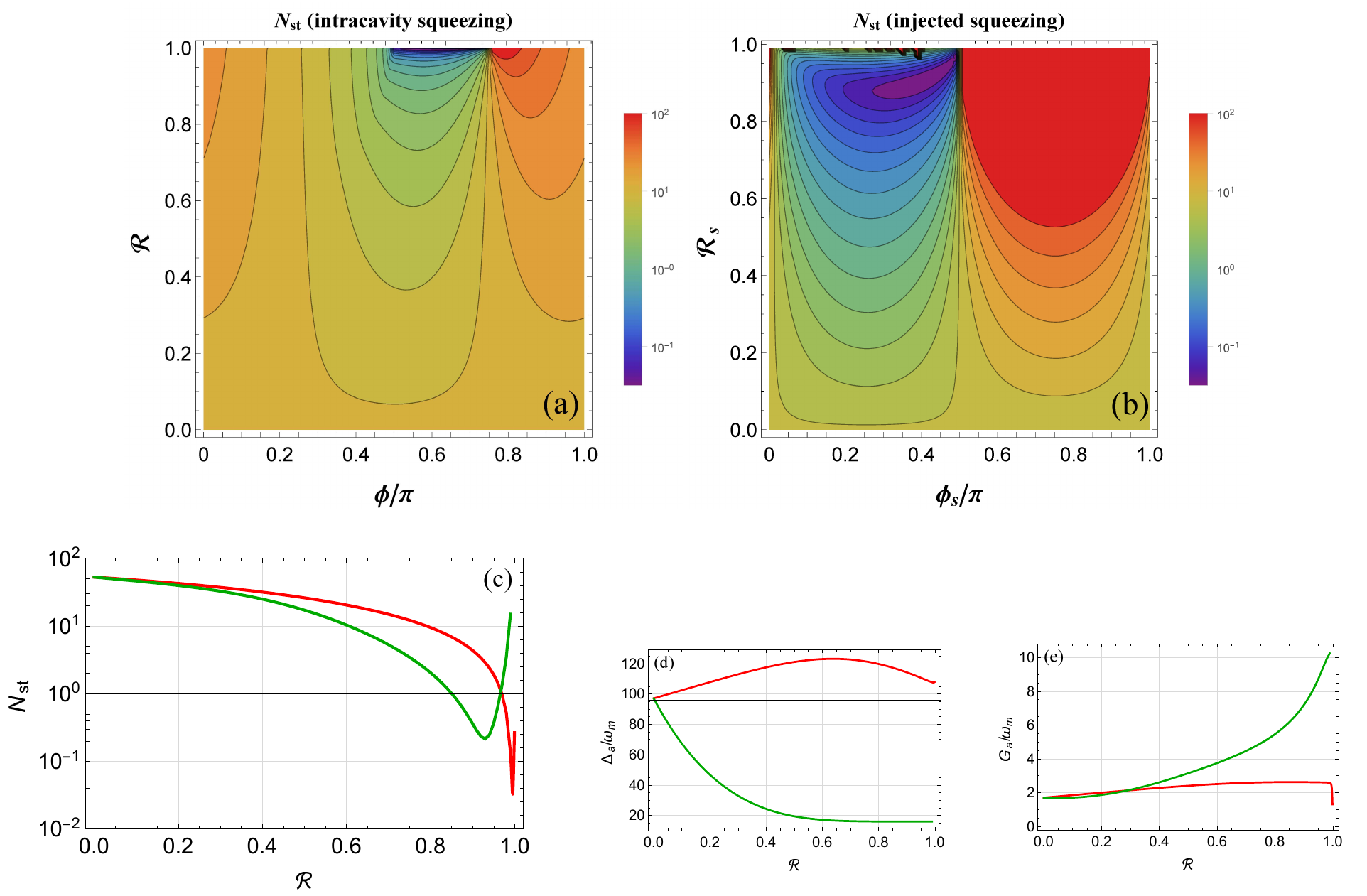}
 \caption{The same as in Fig.~\rp{fig4}, but with $\kappa_a=100\omega_m$.  The curves in (c), (d) and (e) correspond to $\phi=0.63\pi$ and  $\phi_s=0.45\pi$.
}\label{fig5}
\end{figure*}

\subsection{Cooling efficiency}
\label{cool1}

We observe that, when Stokes processes are suppressed, the anti-Stokes scattering rate reduces to
\begin{eqnarray}
A_{a,-}\Bigl|_{\phi=\phi\al{opt},\ \chi=\chi\al{opt}}
=\frac{2\,\kappa_a\,G_a^2}{\kappa_a^2+\pt{\Delta_a-\omega_m}^2}
\end{eqnarray}
in the case of internal squeezing, and this is equal to the value of the standard result for the anti-Stokes rate with no squeezing.
Instead, as discussed in Ref.~\cite{Asjad2016a}, in the case of injected squeezing, the value of the rate for anti-Stokes processes is lowered according to the relation
\begin{eqnarray}
A_{a,-}\al{s}\Bigl|_{\phi_s=\phi_s\al{opt},\ n_s=n_s\al{opt}}
=\frac{2\,\kappa_a\,{G_a\al{s}}^2}{\kappa_a^2+\pt{{\Delta_a\al{s}}-\omega_m}^2}-\frac{2\,\kappa_a\,{G_a\al{s}}^2}{\kappa_a^2+\pt{{\Delta_a\al{s}}+\omega_m}^2}\ .\nn\\
\end{eqnarray}
In this case, in fact, the optical cooling rate $\Gamma_a\al{s}=A_{a,-}\al{s}-A_{a,+}\al{s}$ remains unchanged with respect to standard sideband cooling~\cite{Asjad2016a}.

Let us, now analyze the cooling performance for the two models at equal value of the optomechanical coupling strength $G_a=G_a\al{s}$. Specifically we are interested in the regime of Stokes scattering suppression defined by Eqs.~\rp{optOPO} and \rp{optSQ}. Under these conditions, Eq.~\rp{chi2s} and the relation $G_a=G_a\al{s}$ imply
\begin{eqnarray}
\Delta_a\al{s}&=&\omega_m
\nn\\
\Delta_a&=&\frac{\kappa_a^2+2\,\omega_m^2}{2\,\omega_m}\ ,
\end{eqnarray}
meaning that for equal optomechanical coupling strength, and under the conditions of Stokes scattering suppression, the two models can be mapped one into the other (i.e. the two models are equivalent) only when $\Delta_a\al{s}=\omega_m$.
As discussed in Ref.~\cite{Asjad2016a}, $\Delta_a\al{s}=\omega_m$, in fact, is the condition of optimal cooling
for the case of external squeezing and sufficiently small $\kappa_a$ and $G_a\al{s}$.
Hence we analyze this situation
in Figs.~\ref{fig2} and \ref{fig3}
where we compare the results with the internal (red lines) and external (green lines) squeezing. In these figures we also plot, as a reference, the corresponding results achievable with standard sideband cooling (gray lines). We report the results as a function of the squeezing phase $\phi=\phi_s$, the value of squeezing $\RR=\RR_s$, the detuning $\Delta_a=\Delta_a\al{s}$ and the optomechanical coupling $G_a=G_a\al{s}$. Each line is evaluated for the parameters that optimize the cooling as reported in the other plots on the same row.  In particular, the plots as a function of $G_a$ and $\Delta_a$  (second and third columns) are evaluated for the values of $\RR$ and $\phi$ which result in Stokes scattering suppression (that correspond to the minima in the plots as a function of $\phi$ and $\RR$, corresponding to the first and fourth columns).
Moreover the plots as a function of $G_a$ are evaluated for $\Delta_a\al{s}=\omega_m$.
As discussed above, under these conditions, the two models access the regime of Stokes scattering suppression (and cooling is optimized) when $G_a=G_a\al{s}$. Correspondingly, we observe in the plot as a function of the optomechanical coupling (third column) that the results for the two models are equal. We also note that, in general, the two models are equivalent in terms of the minimum number of mechanical excitations; however, in the case of internal squeezing, larger squeezing is needed in order to achieve comparable results.
Fig.~\ref{fig2} shows the cooling performance for three different values of the cavity decay rate, corresponding to the three rows. From top to bottom the value of $\kappa_a$ increases up to values far beyond the resolved sideband limit.
At small $\kappa_a$ the improvement of the cooling efficiency due to squeezed light is very small. Significant enhancement is instead observed when the system operates not in the resolved sideband regime. In this case ground state cooling with surprisingly high fidelity is achieved even for large values of the cavity decay rate $\kappa_a\gg\omega_m$.
Fig~\ref{fig3} shows, instead, the results for different temperatures when $\kappa_a$ is equal to the one used in the second row  of Fig.~\ref{fig2}. Specifically first and second rows are evaluated for lower and higher temperature respectively. As expected we observe that the enhanced efficiency due to squeezed light is in general observable for sufficiently low temperature
that is, for example, achievable in a cryostat. For example, in the case of a resonator with $\omega_m=2\pi\times 2\,$MHz, the results in Fig.~\ref{fig2}, would correspond to a temperature of $\sim\, 100\,$mK, while the second row of Fig.~\ref{fig2} to $\sim\,10$K.

\subsection{Ground state cooling beyond the resolved sideband regime}
\label{cool2}

Let us now look more closely to the regime of large $\kappa_a$ (third row of Fig.~\ref{fig2}). Typically in this regime standard sideband cooling is not efficient. However we observe that the resonator is cooled to the ground state with high fidelity even if the corresponding sideband cooling result would predict a number of excitations larger than $10$.
This is observed for both models; however
in the case of internal squeezing the optimal parameter choice occurs close to the system instability as observed in Fig.~\rp{fig2} (j), where stability occurs only in a relatively narrow range of the squeezing phase $\phi$.
We also observe that in this case ground state cooling is achieved
for an optomechanical coupling larger than the mechanical frequency. This result hence occurs beyond the regime of validity of the analytical result discussed in \ref{secStokes}.
Hence, the study that we have discussed in the previous section, which focuses onto the regime of Stokes scattering suppression, is too limited in this case. Furthermore while in  Fig.~\ref{fig2} we have constrained the system parameters within the regime in which the two models are equivalent $\chi<\abs{\Delta_a}$ (i.e. $\RR<\abs{\Delta_a}/\sqrt{\kappa_a^2+\Delta_a^2}$), here we actually find that for large $\kappa_a$ and in the case of internal OPO, the optimal cooling is obtained for values of $\chi$ larger than $\abs{\Delta}$ (i.e. for $\RR>\abs{\Delta}/\sqrt{\kappa_a^2+\Delta_a^2}$). This is reported in Figs.~\ref{fig4} and \ref{fig5},
where we explore this regime in more detail. In plots (a) and (b) of both
Fig.~\ref{fig4} and \ref{fig5} we report the steady state phonon number as a function of the squeezing phase $\phi$ ($\phi_s$) and of the value of squeezing $\RR$ ($\RR_s$) for both models and large values of $\kappa_a$, namely $\kappa_a=10\omega_m$ in Fig.~\ref{fig4} and $\kappa_a=100\omega_m$ in Fig.~\ref{fig5}.
Cuts along specific values of the phases $\phi$ and $\phi_s$ corresponding to the minimum $N_{st}$ are reported in the plots (c) of the two figures.
These plots are evaluated minimizing the value of $N_{st}$, for each point in the $\phi-\RR$ space, as a function of both $\Delta_a$ and $G_a$. The specific values of $\Delta_a$ and $G_a$ corresponding to the plots (c) are reported in plots (d) and (e) of Figs.~\ref{fig4} and \ref{fig5}.
In the case of internal OPO, the minimum of $N_{st}$ is observed at large squeezing $\RR\sim1$.
Specifically, the larger is the cavity linewidth, the larger is the value of $\RR$.
In the case of external squeezing, instead, the minimum of $N_{st}$ is achieved at lower values $\RR$ and the system results more stable [the dip in plot (c) has a larger width].
We note that, while in both cases the resonator is cooled to the ground state with high fidelity, lower mechanical energy is observed in the case of internal squeezing. In particular, while in the case of injected squeezing the cooling efficiency is reduced as $\kappa_a$ increases, the value of the minimum of $N_{st}$ with internal squeezing is essentially unaffected by the specific value of the cavity linewidth.

\section{Effects of the OPO pump field}
\label{Sec:OPOpump}

\begin{figure*}[th!]
\centering
 \includegraphics[width=.8\textwidth]{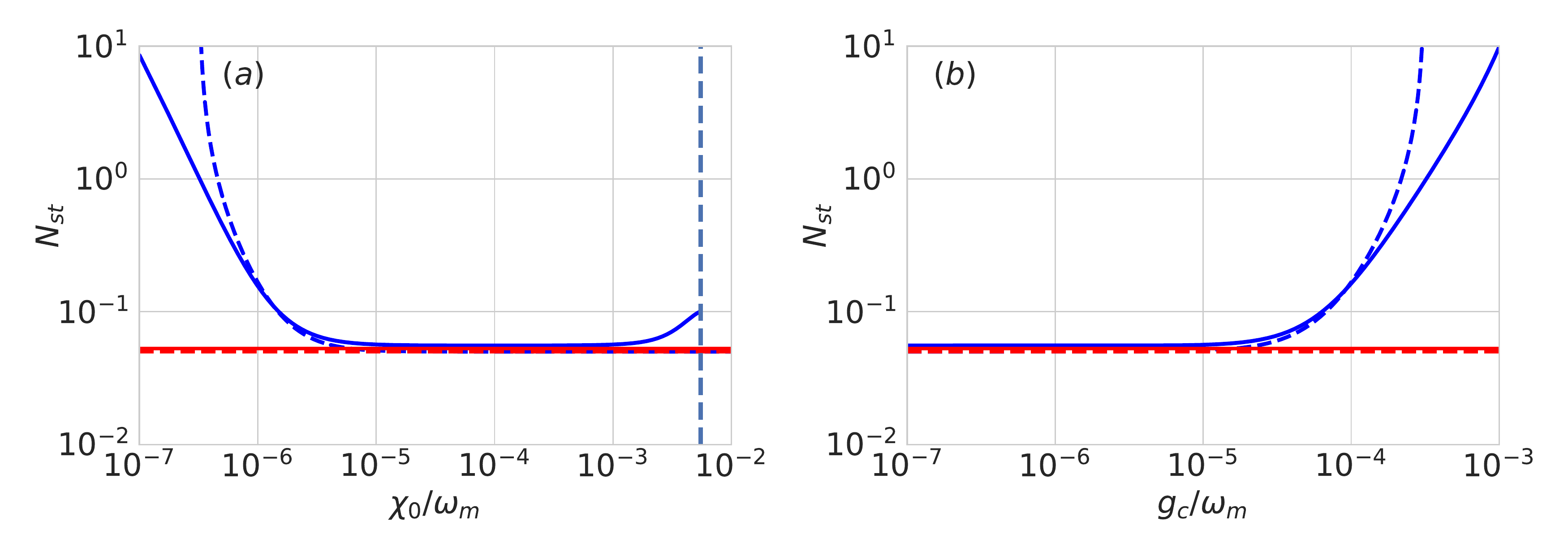}
 \caption{Effect of the OPO pump. Steady state excitation number as a function of (a) the single photon non-linearity $\chi_0$ and (b) single photon optomechanical coupling strength $g_c$ for the pump mode, for fixed values of $G_a$, $\chi$ and $g_a$.
Solid blue lines are numerical results evaluated solving Eq.~\rp{OPO1}, dashed blue lines are evaluated using the analytical result in Eq.~\rp{Apmtot}. The red solid and dashed red lines are for the model in Eq.~\rp{OPO0} and the corresponding approximations in Eq.~\rp{Sa}. The dashed vertical line limits the region of stability of the model in Eq.~\rp{OPO1}.
The solid blue lines are evaluated for values of the cavity linewidth and the optical detuning, corresponding to mode $a$, modified
according to the relations $\kappa_a-{\kappa_c\ \epsilon^2}\pt{\kappa_c^2+\Delta_c^2}$ and $\Delta_a+{\Delta_c\ \epsilon^2}/\pt{\kappa_c^2+\Delta_c^2}$, where $\kappa_a=\Delta_a=\omega_m$ are the corresponding values used for the red lines.
All the results are evaluated for $g_a=10^{-6}\,\omega_m$, $\kappa_c=500\omega_m$ and $\kappa_c=\omega_m$ and for the parameters that minimize $N_{st}$ in the second row of Fig~\ref{fig2}. In (a) $g_c=10^{-6}\omega_m$ and in (b) $\chi_0=10^{-4}\omega_m$.
 }\label{fig6}
 \end{figure*}

In the case of internal squeezing also the OPO pump field interacts with the mechanical resonator.  Therefore, in this case, it is important to analyze its effect on the cooling dynamics.
In general we expect that the results of the previous section are valid as far as $\abs{\kappa_c+\ii\Delta_c}$ is sufficiently large. Here we analyze what are the dominant effects when this condition is not exactly fulfilled.
We can extend the approximate analytical result presented in the previous section
by including also the mechanical effects of the pump at the lowest relevant order in the optomechanical coupling $G_c$ and in the light-modes coupling $\epsilon$.
Specifically, we can decompose the scattering rates, defined in Eq.~\rp{fullApm}, as the sum of three contributions
\begin{eqnarray}\label{Apmtot}
A_\pm=
G_a^2\ S_a(\mp\omega_m)+G_c^2\ S_c(\mp\omega_m)+G_a\,G_c\ S_{a,c}(\mp\omega_m),
\end{eqnarray}
where the first is due to the mode $a$ and is equal to the one reported in Eq.~\rp{Sa}, the second is due to the OPO pump mode $c$, and the last is due to the cross correlations. They are defined by the relations $S_c(\omega)=\int_{-\infty}^\infty\,\dd t\ \ee^{\ii\omega\,t}\av{X_c(t)\ X_c(0)}_{st}$, 
with $X_c=\delta c+\delta c\da$, and $S_{a,c}(\omega)=\int_{-\infty}^\infty\,\dd t\ \ee^{\ii\omega\,t}
\av{X_c(t)\ X_a(0)+X_a(t)\ X_c(0)}_{st}$, 
where the averages are evaluated for a cavity with no mechanical resonator using Eq.~\rp{OPO1} and expanding the result at first order in $\epsilon$.
The explicit expressions for the spectra are
\begin{eqnarray}
S_c(\omega)&=&\frac{2\,\kappa_c}{\kappa_c^2+\pt{\Delta_c-\omega}^2},
\nn\\
S_{a,c}(\omega)&=&\frac{2\,\kappa_a\ \epsilon}{\abs{\zeta(\omega)}^2}{\rm Re}\lpg{
\nu(\omega)\ \ee^{-2\ii\phi}\lpq{
\frac{\kappa_c}{\kappa_a}\frac{\zeta(\omega)}{\kappa_c^2+\pt{\Delta_c-\omega}^2}
}
}\nn\\&&\rpg{\rpq{
-\frac{\kappa_a-\ii\pt{\Delta_a+\omega}}{\kappa_c+\ii\pt{\Delta_c-\omega}}
-\frac{\chi\ \ee^{2\ii\phi}}{\kappa_c-\ii\pt{\Delta_c+\omega}},
}
}
\end{eqnarray}
where $\nu(\omega)$ and $\zeta(\omega)$ are defined in Eq.\rp{nuzeta}.
We note that when Stokes scattering is suppressed $\nu(-\omega_m)=0$, so that also the contribution to the Stokes rate $A_+$ due to the cross correlation is zero, i.e. $S_{a,c}(-\omega_m)=0$. The contribution due to the OPO pump $S_c\pt{-\omega_m}$, instead, remains and it is responsible for the enhancement of the Stokes scattering, so that it inevitably reduces the cooling efficiency. Therefore at lowest order the contribution of the OPO pump is to reduce the cooling efficiency, and optimal cooling is achieved
when
the mechanical effect of the pump field is negligible.

We have tested this prediction by comparing the results obtained form the model in Eq.~\rp{OPO0} with those obtained by solving numerically Eq.~\rp{OPO1}.
Specifically, we consider parameters consistent with realistic lithium-niobate microdisk cavities which can exhibit large non-linearity and constitute the OPO~\cite{Furst2010,Fortsch,Furst,Otterpohl}. The resonator can be instead either a vibrational mode of the microdisk itself~\cite{Hofer}, or of a nearby evanescently coupled mechanical resonator~\cite{Anetsberger}.
In these systems the single photon non-linearities ($\chi_0$) are of the order of kHz while the single photon optomechanical interaction strength ($g_a$) for a MHz mechanical mode (for example the breathing mode), is below one Hz~\cite{Peano}.

In our comparison, reported in Fig.~\ref{fig6}, we fix the optimal values of $G_a$ and $\chi$ for the system parameters used in the second row of Fig.~\ref{fig2},
and analyze the cooling performance as a function of $g_c$ and $\chi_0$, so that
$G_2=g_b\ \chi/\chi_0$ and $\epsilon=\chi_0\ G_a/g_a$.
In general the mechanical effects of mode c are negligible for small $g_c$ as described in Fig.~\ref{fig6} (b), where the result for the full model in Eq.~\rp{OPO1} (blue lines) reproduce that of model in Eq.~\rp{OPO0} (red lines) for sufficiently small $g_c$.
Moreover, on the one hand, if $\chi_0$ is too small then a strong pump is required to achieve sufficiently large $\chi$ so that also $G_c$ is large (while $\epsilon$ is small), and the mechanical effect of mode c is significant so
that the model discussed in the previous section may fail. On the other hand, if $\chi_0$  is large then $G_c$ is small but $\epsilon$ is large, and this strong optical coupling can spoil the dynamics discussed in the previous section.
Hence we expect that the model described by Eq.~\rp{OPO0} is valid for intermediate values of $\chi_0$.
This behaviour is described by Fig.~\ref{fig6} (a). In the limit of large $\chi_0$, which imply small $G_c$, and hence negligible direct mechanical effect, the effect of mode coupling $\epsilon$, that can be instead significant, can be taken into account at lowest order in $1/\abs{\kappa_c+\ii\Delta_c}$ by considering renormalized cavity detuning $\Delta_a$ and linewidth $\kappa_a$.
In details, in Fig.~\ref{fig6} we fix the values of the parameters of the reduced model~\rp{OPO0}, so that the blue curves, corresponding to the full model~\rp{OPO1}, have been evaluated by including the modification due to the mode coupling according to the relations $\Delta_a\to\Delta_a+{\Delta_c\ \epsilon^2}/\pt{\kappa_c^2+\Delta_c^2}$ and
$\kappa_a\to\kappa_a-{\kappa_c\ \epsilon^2}\pt{\kappa_c^2+\Delta_c^2}$.

\section{Conclusions}\label{conclusions}

In conclusion we have studied how optomechanical sideband cooling is modified when the fluctuations of the cooling light are squeezed by a degenerate optical parametric oscillator (OPO). We have considered the situation in which the non-linear medium which provides the parametric interaction is placed inside the optomechanical system and we have shown that in specific limits this system is equivalent to a system in which
the light generated by the OPO is injected into the optomechanical system. It has been demonstrated~\cite{Asjad2016a,Clark2017} for the latter that Stokes scattering transitions can be suppressed and the cooling efficiency enhanced. Correspondingly we show that an equal suppression of the Stokes processes can be observed also in the case of internal OPO, so that also in this case the backaction limit can be surpassed.
However, although the two systems are ideally equivalent, they achieve equal results under different physical conditions,
and each of them exhibit specific advantages and disadvantages.
On the one hand the internal OPO scheme has the advantage of bypassing the detrimental effects of the inevitable losses in the transmission and injection of the squeezed light from the external OPO into the optomechanical cavity. On the other hand it is negatively affected by the OPO pump field which can also interact with the mechanical resonator and reduce the cooling efficiency. Furthermore, the system with the internal OPO requires higher values of squeezing and is optimal close to the instability of the system.

A further important advantage of adding squeezing within an optomechanical system is that it allows to achieve ground state cooling in the unresolved sideband regime, characterized by large cavity decay rate $\kappa_a$, which would be otherwise impossible.
This result is observed for both intracavity and injected squeezing
but, in this regime, optimal cooling is observed in a parameter regime in which the two models are not equivalent; in this case the scheme with internal squeezing results clearly superior, although it requires a precise fine-tuning of the squeezing strength close to the optical instability threshold.

We finally point out that the present analysis applies also to other, apparently different, physical systems.
One can, for example, consider an hybrid configuration in which a Fabry-Perot optical cavity interacts with both a mechanical resonator and an array of two or more transversally driven two-level atoms~\cite{Begley,Neuzner} which can realize the parametric two-photon driving~\cite{Fernandez-Vidal,Habibian}. The intracavity atoms provide the parametric nonlinearity, with the advantage of avoiding the detrimental effects of the OPO pump. Another possibility could be to include a Josephson parametric amplifier within the microwave cavity of an optomechanical system similar to that discussed in Refs.~\cite{Clark}.


\section*{Acknowledgments}

\end{document}